\documentclass[aps,pra,preprintnumbers,showpacs,tightenlines]{revtex4}
\usepackage{amssymb}
\usepackage{amsmath}
\usepackage{graphicx}
\usepackage{epsfig}
\usepackage{subfigure}
\usepackage{amsfonts}
\usepackage{CJK}

\begin{document}

\title{Generating double NOON states of photons in circuit QED}
\author{Qi-Ping Su$^{1}$, Hui-Hao Zhu$^{1}$, Li Yu$^{1}$, Yu Zhang$^{1}$, Shao-Jie Xiong$^{2}$, Jin-Ming Liu$^{2}$,}
\author{Chui-Ping Yang$^{1}$}
\email{yangcp@hznu.edu.cn}
\address{$^1$Department of Physics, Hangzhou Normal University, Hangzhou, Zhejiang 310036, China}
\address{$^2$State Key Laboratory of Precision Spectroscopy, Department of Physics, East China Normal University, Shanghai
200062, China}
\date{\today}

\begin{abstract}
To generate a NOON state with a large photon number $N$, the number of operational steps could be large
and the fidelity will decrease rapidly with $N$. Here we propose a method to generate a new type of quantum entangled states,
$(|NN00\rangle+|00NN\rangle)/\sqrt{2}$ called ``double NOON" states, with
a setup of two superconducting flux qutrits and five circuit cavities.
This scheme operates essentially by employing a two-photon process, i.e., two
photons are simultaneously and separately emitted into two cavities when
each coupler qutrit is initially in a higher-energy excited state. As a
consequence, the ``double" NOON state creation needs only $N$+2 operational steps. One application of double NOON states is to get a phase error of $1/(2N)$
in phase measurement. In comparison, to achieve the same error, a normal NOON state of the form $(|2N,0\rangle+|0,2N\rangle)/\sqrt{2}$ is needed,  which requires at least $2N$ operational steps to prepare by using the existing schemes. Our numerical simulation demonstrates that high-fidelity generation
of the double NOON states with $N\leq 10$ even for the imperfect devices is
feasible with the present circuit QED technique.
\end{abstract}

\maketitle
\date{\today}

\begin{center}
\textbf{I. INTRODUCTION}
\end{center}

Superconducting qubits have attracted substantial attention because of their
controllability, integrability, ready fabrication and potential scalability
\cite{s01,s02,s03,s04} in quantum information and quantum computation. The
level spacings of superconducting qubits can be rapidly adjusted ($1\sim3$
ns) by varying external control parameters \cite{s05,s06,s07,s08}. Their
coherence time is increasing rapidly \cite{s081,s082,s083,s084,s085,s086}.
The strong coupling and ultrastrong coupling of one qubit with one microwave
cavity have been reported in experiments \cite{s09,s00}. The circuit QED is
considered as one of the most feasible candidates for quantum computation
\cite{s03,s04}.

The NOON states, $(|N0\rangle _{12}+|0N\rangle _{12})/\sqrt{2}$ with $N$
photons in mode 1 or 2, have attracted considerable attention because of
their significant applications in quantum optical lithography \cite{s1,s4},
quantum metrology \cite{s2,s201,s202}, precision measurement \cite{s3}, and
quantum information processing \cite{s31}. For example, if a photon in mode
2 \textrm{gains} an extra phase $e^{i\phi }$ compared with a photon in mode
1, the NOON state of $N$ photons will become $(|N0\rangle _{12}+e^{iN\phi
}|0N\rangle _{12})/\sqrt{2}$ (a common phase has been ignored). With this
state, the measurement of $A_{N}=(|N0\rangle _{12}+|0N\rangle _{12})(\langle
N0|_{12}+\langle 0N|_{12})/2$ gives $\langle A_{N}\rangle =[1+\cos {(N\phi )}%
]/2$. Note that $A_{N}^{2}=A_{N}$, one has $\left( \Delta A_{N}\right)
^{2}=\langle A_{N}^{2}\rangle -\langle A_{N}\rangle ^{2}=\sin ^{2}{(N\phi )/4%
}$. According to \cite{s4}, the error in the phase is calculated as:
\begin{equation}
\Delta \phi =\Delta A_{N}/\left\vert d\langle A_{N}\rangle /d\phi
\right\vert =1/N,
\end{equation}%
which reaches the Heisenberg limit $1/\bar{N}$ ($\bar{N}$ is the average
number of photons counted in a chosen time interval) \cite{s41}. It is
obvious that the error of phase measurement will be better for a larger
photon number $N$, but the high-fidelity generation of NOON states with a
large $N$ is not easy in experiments.

Some schemes have been presented for the generation of the NOON states, $%
(|N0\rangle +|0N\rangle )/\sqrt{2},$ of photons in two cavities or
resonators. The setup in Ref. \cite{s8} consists of two superconducting
cavities and a tunable qubit, and the qubit alternatively interacts with two
cavity modes and a classical pulse respectively. According to [25],
generating a NOON state requires a linear number of operations, which is
greater than $N$ (e.g., 12 basic operations required for creating a NOON
state with $N=3$). The circuit in Ref. \cite{s9} is more complicated, which
requires three superconducting resonators and two qutrits, but only $N+1$
operational steps are needed. The scheme presented in Ref. \cite{s9} has
been implemented in experiments to generate a NOON state with $N\leq 3$ \cite%
{s5}. Ref. \cite{s7} adopts a setup consisting of one superconducting
transmon qutrit and two resonators, which is simpler than that in Refs. \cite%
{s9,s5}. Though $2N$ steps of operation are required to generate a NOON
state with $N$ photons, the generation of NOON states in this scheme is
faster than that in Ref. \cite{s8}. Then, the scheme in Ref. \cite{s7} is
improved in Ref. \cite{s13}, in which one four-level superconducting flux
device and two resonators are required while only $N+1$ operational steps
are needed. From the discussion given here, one can see that by using
schemes [25-29] to generate a NOON state $(|N0\rangle +|0N\rangle )/\sqrt{2}$
of photons, at least $N$ operational steps are needed.

In this paper, we propose an efficient scheme for generating the
\textquotedblleft double\textquotedblright\ NOON states $(|NN00\rangle
+|00NN\rangle )/\sqrt{2}$ in four cavities or resonators with only $N+2$
operational steps (including two basic steps for initial preparation of a
two-qutrit Bell state). The setup consists of two superconducting flux
qutrits and five cavities (see Fig. 2). A two-photon process is adopted in
this scheme, which is quite different from the single-photon process used in
Ref. \cite{s8,s9,s5,s7,s13}. It is the first time to demonstrate that a
double NOON state, a new type of NOON states of photons, can be generated in
cavity/circuit QED.

For the concrete use of the double NOON states, let us consider their
application in phase measurement, which can be implemented by using a setup
illustrated in Appendix. Assume that photons in four modes ($1,2,3,4$)
are initially in a double NOON state $(|NN00\rangle _{1234}+|00NN\rangle
_{1234})/\sqrt{2}.$ Each photon in mode $3$ ($4$) experiences an extra phase
shift $\phi $ compared to each photon in mode $1$ ($2$), which is induced by
a phase shifter. When the photons reach their detectors, the double NOON
state of photons in four modes ($1,2,3,4$) can thus be expressed as  $|\Psi
\rangle =(|NN00\rangle _{1234}+e^{i2N\phi }|00NN\rangle _{1234})/\sqrt{2}$
(a common phase has been ignored). The measurement of $A_{DN}=(|NN00\rangle
_{1234}+|00NN\rangle _{1234})(\langle NN00|_{1234}+\langle 00NN|_{1234})/2$
gives:
\begin{equation}
\langle A_{DN}\rangle =\langle \Psi |\cdot A_{DN}\cdot |\Psi \rangle
=[1+\cos (2N\phi )]/2~.
\end{equation}%
In addition, one has $\left( \Delta A_{DN}\right) ^{2}=\frac{1}{4}\sin ^{2}{%
(2N\phi ).}$ Thus, the error of phase measurement will be $\Delta \phi
=1/(2N)$, which also reaches the Heisenberg limit. In comparison, to achieve
the same error for phase measurement, a normal NOON state of the form $%
(|2N,0\rangle +|0,2N\rangle )/\sqrt{2}$ is needed, which requires at least $%
2N$ operational steps to prepare by using the existing schemes [25-29].

This paper is arranged as follows. In Sec. II, we introduce the effective
Hamiltonian and time evolution for the left-hand half as well as the
right-hand half of the setup [Fig. 2(a)]. In Sec. III, we describe how to
generate the double NOON states of photons in four cavities or resonators.
In Sec. IV, we discuss the possible experimental implementation of this
proposal.

\begin{center}
\textbf{II. EFFECTIVE HAMILTONIAN AND TIME EVOLUTION}
\end{center}

The setup, shown in Fig.~1(a), consists of two superconducting flux qutrits
and five cavities. In Fig.~1(a), the left-hand qutrit is labelled as qutrit
\textit{L} while the right-hand qutrit is labelled as qutrit \textit{R.}
Qutrit \textit{L} is coupled to cavities 1 and 2 while qutrit \textit{R }is
coupled to cavities 3 and 4. In addition, both qutrits \textit{L} and
\textit{R} are connected by a common central cavity. Each qutrit has three
energy levels $\left\vert g\right\rangle ,$ $\left\vert e\right\rangle ,$
and $\left\vert f\right\rangle .$ In the following, we first introduce the
effective Hamiltonian and time evolution for the left-hand half of the setup
(i.e., qutrit \textit{L} and cavities 1 and 2). We then introduce the
effective Hamiltonian and time evolution for the right-hand half of the
setup (i.e., qutrit \textit{R} and cavities 3 and 4).

\begin{figure}[tbp]
\begin{center}
\includegraphics[bb=0 0 1220 790, width=9.5 cm, clip]{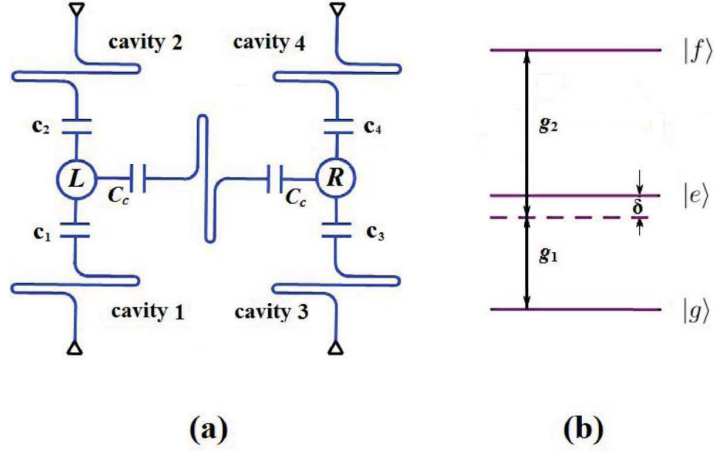} \vspace*{%
-0.08in}
\end{center}
\caption{(Color online) (a) Setup for two superconducting flux qutrits and
five cavities. A common cavity (i.e., the central one) is used to initially
prepare the two qutrits in a Bell state, while the other four cavities are
utilized to create the double NOON state. Here, each cavity is a
one-dimensional transmission line resonator. The left-hand circle represents
the qutrit labled by \textit{L}; while the right-hand circle represents the
qutrit labeled by \textit{R}. (b) Illustration of each qutrit dispersively
interacting with its two cavities. Here, $g_1$ is the coupling strength
between cavity 1 (3) with the $|g\rangle\leftrightarrow|e\rangle$ while $g_2$
is the coupling strength between cavity 2 (4) with the $|e\rangle%
\leftrightarrow|f\rangle$ transition. In addition, we set $\delta_1=\delta_2=\delta$.}
\label{fig:1}
\end{figure}

Suppose that cavity 1 (2) is coupled to the $\left\vert g\right\rangle
\leftrightarrow \left\vert e\right\rangle $ ($\left\vert e\right\rangle
\leftrightarrow \left\vert f\right\rangle $) transition of the coupler
qutrit $L$. In the interaction picture, the Hamiltonian describing the
interaction between qutrit $L$ and the two cavities 1 and 2 is (assuming $%
\hbar =1$)
\begin{equation}
H_{I}(a_{1},a_{2})=g_{1}(e^{i\delta _{1}t}a_{1}\sigma
_{ge,L}^{+}+h.c.)+g_{2}(e^{-i\delta _{2}t}a_{2}\sigma _{ef,L}^{+}+h.c.),
\end{equation}%
where $\delta _{1}=\omega _{ge,L}-\omega _{c1}>0$, $\delta _{2}=\omega
_{c2}-\omega _{ef,L}>0$, $\omega _{ge}$ ($\omega _{ef}$) is the $\left\vert
g\right\rangle \leftrightarrow \left\vert e\right\rangle $ ($\left\vert
e\right\rangle \leftrightarrow \left\vert f\right\rangle $) transition
frequency, $\omega _{c1}$ ($\omega _{c2}$) is the frequency of cavity $1$ ($2
$), and $g_{1}$ ($g_{2}$) is the coupling strength between cavity $1$ ($2$)
and the $\left\vert g\right\rangle \leftrightarrow \left\vert e\right\rangle
$ ($\left\vert e\right\rangle \leftrightarrow \left\vert f\right\rangle $)
transition [Fig.~1(b)]. In addition, $\sigma _{ge,L}^{+}=|e\rangle _{L}\langle g|$, $%
\sigma _{ef,L}^{+}=|f\rangle _{L}\langle e|$, and $a_{1}$ $(a_{2})$ is the
photon annihilation operator of cavity $1$ ($2$).

Under the large-detuning condition $\delta_1>>g_{1}$ and $\delta_2>>g_{2}$, the dynamics governed by $H_{I}(a_{1},a_{2})$ is equivalent
to that decided by the following effective Hamiltonian \cite{s14}
\begin{equation}
H_{eff}=H_{0}+H_{int},
\end{equation}%
with~~~~~~~~~~~~~~~~~~
\begin{eqnarray}
H_{0} &=&-\frac{g_{1}^{2}}{\delta _{1}}(a_{1}^{\dagger }a_{1}|g\rangle
_{L}\langle g|-a_{1}a_{1}^{\dagger }|e\rangle _{L}\langle e|)-\frac{g_{2}^{2}%
}{\delta _{2}}(a_{2}a_{2}^{\dagger }|f\rangle _{L}\langle f|-a_{2}^{\dagger
}a_{2}|e\rangle _{L}\langle e|), \\
H_{int} &=&-\lambda \left[ e^{-i\left( \delta _{1}-\delta _{2}\right)
t}a_{1}^{\dagger }a_{2}^{\dagger }|g\rangle _{L}\langle f|+h.c.\right] ,
\end{eqnarray}%
where $\lambda =\frac{g_{1}g_{2}}{2}\left( \frac{1}{\delta _{1}}+\frac{1}{%
\delta _{2}}\right) .$ In a new interaction picture with respect to the free
Hamiltonian $H_{0}$, one has
\begin{eqnarray}
\widetilde{H}_{int} &=&e^{iH_{0}t}H_{int}e^{-iH_{0}t}  \notag \\
&=&-\lambda \left[ e^{-i\left( \delta _{1}-\delta _{2}\right) t}e^{-i\frac{%
g_{1}^{2}}{\delta _{1}}a_{1}^{+}a_{1}t}a_{1}^{+}a_{2}^{+}|g\rangle
_{L}\langle f|e^{i\frac{g_{2}^{2}}{\delta _{2}}a_{2}a_{2}^{+}t}+h.c.\right] .
\end{eqnarray}

Denote $\left\vert n\right\rangle _{j}$ as the $n$-photon state for cavity $%
j $ ($j=1,2$). Under the Hamiltonian (7), the state $\left\vert
f\right\rangle _{L}\left\vert n\right\rangle _{1}\left\vert n\right\rangle
_{2}$ evolves in a subspace formed by two orthogonal states $\left\vert
f\right\rangle _{L}\left\vert n\right\rangle _{1}\left\vert n\right\rangle
_{2}$ and $\left\vert g\right\rangle _{L}\left\vert n+1\right\rangle
_{1}\left\vert n+1\right\rangle _{2}$. In this subspace, the Hamiltonian (7)
can be expressed as

\begin{equation}
\widetilde{H}_{int}=\left(
\begin{array}{cc}
0 & -e^{i\left( \delta _{1}-\delta _{2}\right) t}e^{i\varphi \left(
n+1\right) t}\left( n+1\right) \lambda  \\
-e^{-i\left( \delta _{1}-\delta _{2}\right) t}e^{-i\varphi \left( n+1\right)
t}(n+1)\lambda  & 0%
\end{array}%
\right) ,
\end{equation}%
where $\varphi =g_{1}^{2}/\delta _{1}-g_{2}^{2}/\delta _{2}.$ To make the
Hamiltonian (8) time independent, the conditions $\delta
_{1}=\delta _{2}$ and $g_{1}=g_{2}$ (i.e., $\varphi=0$) need to be satisfied. Note that $\delta
_{1}=\delta _{2}$ can be met by adjusting the level spacings of the qutrit
or the cavity frequency, and $g_{1}=g_{2}$ can be achieved by a prior design
of the sample with appropriate capacitances $c_{1}$ and $c_{2}$. Under the
conditions $\delta _{1}=\delta _{2}\equiv \delta $ and  $g_{1}=g_{2}\equiv g,
$ the matrix (8) becomes

\begin{equation}
\widetilde{H}_{int}=\left(
\begin{array}{cc}
0 & -\left( n+1\right) \lambda  \\
-\left( n+1\right) \lambda  & 0%
\end{array}%
\right) ,
\end{equation}%
where $\lambda =g^{2}/\delta $ (which will apply below). Under the Hamiltonian
(9), the time evolution of the state $\left\vert f\right\rangle
_{L}\left\vert n\right\rangle _{1}\left\vert n\right\rangle _{2}$ can be
described as
\begin{equation}
\left\vert f\right\rangle _{L}\left\vert n\right\rangle _{1}\left\vert
n\right\rangle _{2}\rightarrow \cos \left[ \left( n+1\right) \lambda t\right]
\left\vert f\right\rangle _{L}\left\vert n\right\rangle _{1}\left\vert
n\right\rangle _{2}+i\sin \left[ \left( n+1\right) \lambda t\right]
\left\vert g\right\rangle _{L}\left\vert n+1\right\rangle _{1}\left\vert
n+1\right\rangle _{2}.
\end{equation}

We now return to the original interaction picture by applying a unitary
transformation $e^{-iH_{0}t}$ to the right side of Eq. (10). It can be found
that in the original interaction picture, the state transformation (10)
becomes
\begin{equation}
\left\vert f\right\rangle _{L}\left\vert n\right\rangle _{1}\left\vert
n\right\rangle _{2}\rightarrow e^{i\left( n+1\right) \lambda t}\left\{ \cos
\left[ \left( n+1\right) \lambda t\right] \left\vert f\right\rangle
_{L}\left\vert n\right\rangle _{1}\left\vert n\right\rangle _{2}+i\sin \left[
\left( n+1\right) \lambda t\right] \left\vert g\right\rangle _{L}\left\vert
n+1\right\rangle _{1}\left\vert n+1\right\rangle _{2}\right\} .
\end{equation}%
On the other hand, one can easily find that under the effective Hamiltonian
(4), the state $|e\rangle |0\rangle |0\rangle $ evolves into
\begin{equation}
|e\rangle _{L}|0\rangle _{1}|0\rangle _{2}\rightarrow e^{-i\lambda
t}|e\rangle _{L}|0\rangle _{1}|0\rangle _{2}.
\end{equation}

Note that the right-hand half of the setup in Fig.~1(a) has the same
configuration as the left-hand half. Therefore, for qutrit $R$ being
identical to qutrit $L$ and cavity $3$ ($4$) identical to cavity $1$ ($2$),
the Hamiltonian describing the right-hand half of Fig. 2(a) would be
\begin{equation}
H_{I}\left( a_{3},a_{4}\right) =\hbar g_{1}(e^{i\delta t}a_{3}\sigma
_{ge,R}^{+}+h.c.)+\hbar g_{2}(e^{-i\delta t}a_{4}\sigma _{ef,R}^{+}+h.c.),
\end{equation}%
where $\sigma _{ge,R}^{+}=|e\rangle _{R}\langle g|,$ $\sigma
_{ef,R}^{+}=|f\rangle _{R}\langle e|$. Note that this Hamiltonian takes the
same form as the Hamiltonian (3) above. Thus, the state evolution for the
coupler qutrit $R$ and cavities $3$ and $4$ would be the same as those given
in Eqs. (11) and (12), with a replacement of the subscripts $L,1,2$ by $%
R,3,4 $. Namely, under the Hamiltonian (13), we have the following state
evolutions%
\begin{equation}
\left\vert f\right\rangle _{R}\left\vert n\right\rangle _{3}\left\vert
n\right\rangle _{4}\rightarrow e^{i\left( n+1\right) \lambda t}\left\{ \cos
\left[ \left( n+1\right) \lambda t\right] \left\vert f\right\rangle
_{R}\left\vert n\right\rangle _{3}\left\vert n\right\rangle _{4}+i\sin \left[
\left( n+1\right) \lambda t\right] \left\vert g\right\rangle _{R}\left\vert
n+1\right\rangle _{3}\left\vert n+1\right\rangle _{4}\right\} ,
\end{equation}%
\begin{equation}
|e\rangle _{R}|0\rangle _{3}|0\rangle _{4}\rightarrow e^{-i\lambda
t}|e\rangle _{R}|0\rangle _{3}|0\rangle _{4}.
\end{equation}

The results (11), (12), (14) and (15) obtained here will be employed for the
preparation of the \textquotedblleft double" NOON state below.

As shown in the next section, the double NOON state preparation employs the
same operations simultaneously performed on the left-hand two cavities and
the right-hand two cavities. Thus, the combined interaction Hamiltonian
would be
\begin{equation}
H_{I,1}=H_{I}(a_{1},a_{2})+H_{I}(a_{3},a_{4}),
\end{equation}%
where $H_{I}(a_{1},a_{2})$ and $H_{I}(a_{3},a_{4})$ commute with each other.

\begin{center}
\textbf{III. PREPARATION OF THE TWO-QUTRIT BELL STATE}
\end{center}

Each qutrit is initially decoupled from its connected cavities, which can be
achieved by a prior adjustment of the qutrit level spacings via varying
external control parameters \cite{s01,s11}. In addition, assume that the
central cavity is initially in a single-photon state and each qutrit is
initially in the ground state $|g\rangle .$

The Bell state of the two qutrits \textit{L} and \textit{R} is generated via
the following operations:

Adjust the level spacings of the qutrits such that the $|g\rangle
\leftrightarrow |e\rangle $ transition of each qutrit is resonant with the
central cavity but each qutrit is decoupled from other cavities. In the
interaction picture (the same picture is used without mentioning hereafter),
the interaction Hamiltonian describing this operation is $%
h_{I}=\sum\limits_{j=L,R}\mu _{j}(a^{\dagger }\sigma _{ge,j}^{-}+h.c.),$
where $a^{\dagger }$ is the photon creation operator for the central cavity,
$\mu _{j}$ is the coupling strength between the central cavity and the $%
|g\rangle \leftrightarrow |e\rangle $ transition of qutrit $j,$ and $\sigma
_{ge,j}^{-}=|g\rangle _{j}\langle e|$ ($j=L,R$). For simplicity, we assume $%
\mu _{1}=\mu _{2}=\mu $, which applies for identical qutrits $L$ and $R$. It
is straightforward to show that the time evolution of the state $|g\rangle
\left\vert g\right\rangle |1\rangle $ under the Hamiltonian $h_{I}$ is
described by $|g\rangle \left\vert g\right\rangle |1\rangle \rightarrow \cos
(\sqrt{2}\mu t)|g\rangle |g\rangle \left\vert 1\right\rangle -\frac{i}{\sqrt{%
2}}\sin (\sqrt{2}\mu t)(|e\rangle |g\rangle +|g\rangle |e\rangle )|0\rangle $%
. Here, $\left\vert 0\right\rangle $ and $\left\vert 1\right\rangle $ are
the vacuum state and the single-photon state of the central cavity,
respectively. In addition, $|k\rangle |l\rangle \equiv |k\rangle
_{L}|l\rangle _{R}$ with $k,l\in \{g,e,f\}$ (Here and below, for simplicity
we omit the subscripts $L$ and $R$)$.$ One can see that after an interaction
time $\tau =\pi /\left( 2\sqrt{2}\mu \right) ,$ the initial state $|g\rangle
|g\rangle |1\rangle $ of the two qutrits plus the central cavity evolves to
\begin{equation}
\frac{1}{\sqrt{2}}(|g\rangle |e\rangle +|e\rangle |g\rangle )|0\rangle ,
\end{equation}%
where a common phase factor $-i$ is omitted. Eq. (17) shows that the two
qutrits are prepared in the Bell state while the central cavity is in the
vacuum state after the operations. The level spacings of the qutrits need to
be adjusted back to the original level configuration such that the qutrits
are decoupled from the central cavity.

Note that during the operations described below for the double NOON state
creation, the central cavity is not involved. Thus, the central cavity can
be dropped off for simplicity.

\begin{center}
\textbf{IV. GENERATION OF THE DOUBLE NOON STATES}
\end{center}

\begin{figure}[tbp]
\begin{center}
\includegraphics[bb=-4 379 2000 1032, width=10.5 cm, clip]{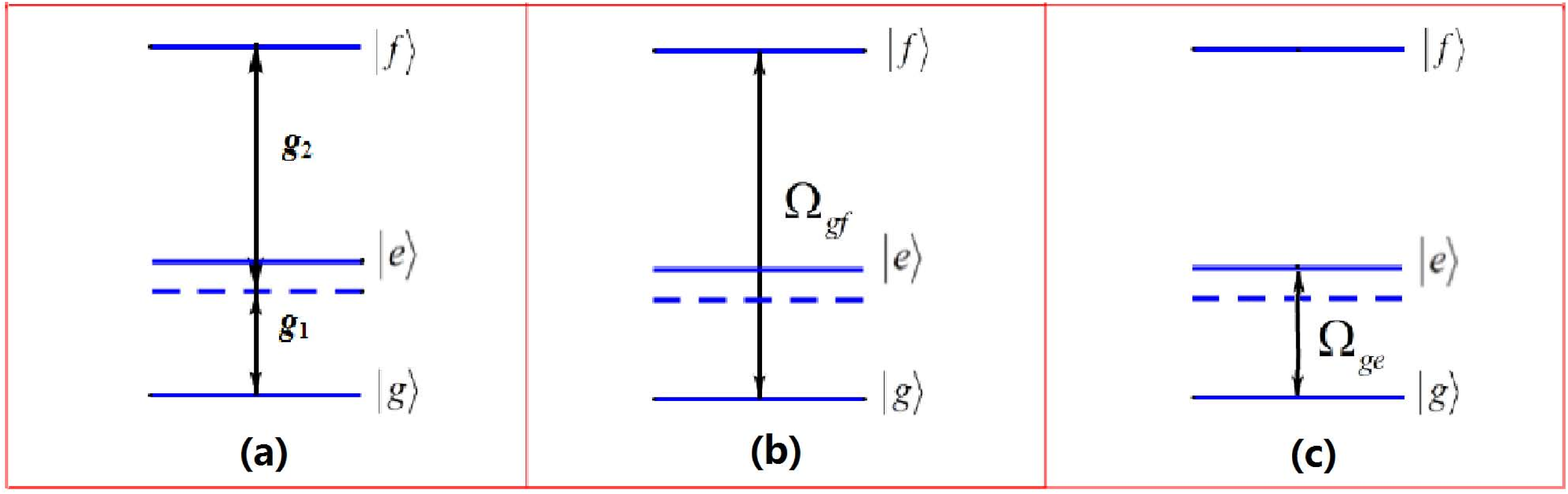}
\vspace*{-0.08in}
\end{center}
\caption{(Color online) (a) Illustration of the dispersive interaction of
qutrit \textit{L} (\textit{R}) with cavities 1 and 2 (3 and 4). $g_{1\text{ }%
}$is the coupling constant between cavity 1 (3) and the $\left\vert
g\right\rangle \leftrightarrow \left\vert e\right\rangle $ transition of
qutrit \textit{L} (\textit{R}), and $g_{2\text{ }}$is the coupling constant
between cavity 2 (4) and the $\left\vert e\right\rangle \leftrightarrow
\left\vert f\right\rangle $ transition of qutrit \textit{L} (\textit{R}).
(b) Illustration of the resonant interaction between the pulse and the $%
\left\vert g\right\rangle \leftrightarrow \left\vert f\right\rangle $
transition of each qutrit with the Rabi frequency $\Omega _{gf}.$ (c)
Illustration of the resonant interaction between the pulse and the $%
\left\vert g\right\rangle \leftrightarrow \left\vert e\right\rangle $
transition of each qutrit with the Rabi frequency $\Omega _{ge}.$ }
\label{fig:2}
\end{figure}

Similar to the NOON state preparation, the double NOON state creation
requires applying classical pulses. For simplicity, we define $\omega _{gf}$
($\omega _{ge}$) as the $|g\rangle \leftrightarrow |f\rangle $ ($|g\rangle
\leftrightarrow|e\rangle $) transition frequency of the qutrits, and define $%
\Omega _{gf}$ ($\Omega _{ge}$) as the Rabi frequency of a classical pulse
driving $|g\rangle \leftrightarrow |f\rangle $ ($|g\rangle \leftrightarrow
|e\rangle $) transition of the qutrits. The frequency, duration, and initial
phase of the pulses are denoted as $\{\omega ,t,\varphi \}$.

Assume that each side cavity is initially in the vacuum state and the two
qutrits are initially prepared in the Bell state. Thus, the state of the two
qutrits and the four side cavities is
\begin{equation}
\frac{1}{\sqrt{2}}(|g\rangle |e\rangle +|e\rangle |g\rangle )|0\rangle
_{1}\left\vert 0\right\rangle _{2}\left\vert 0\right\rangle _{3}\left\vert
0\right\rangle _{4},
\end{equation}%
where the subscripts 1, 2, 3, and 4 indicate the four cavities 1, 2, 3, and
4, respectively.

Before starting the double NOON state preparation, a classical pulse of $\{\omega _{gf},\frac{%
\pi }{2\Omega _{gf}},-\frac{\pi }{2}\}$ needs to be applied to the qutrits
[Fig.~2(b)], which is described by the following Hamiltonian
\begin{equation}
H_{I,2}=\sum\limits_{j=L,R}(\Omega _{{gf}}e^{i\pi /2}\sigma
_{gf,j}^{+}+h.c.),
\end{equation}%
where $\sigma _{gf,j}^{+}=|f\rangle _{j}\langle g|.$ It is easy to find that
under the Hamiltonian (19), we have the state rotation $|g\rangle
\rightarrow \cos \left( \Omega _{{gf}}t\right) \left\vert g\right\rangle
+\sin \left( \Omega _{{gf}}t\right) |f\rangle $, which shows that the pulse
with a duration $\frac{\pi }{2\Omega _{gf}}$ results in $|g\rangle
\rightarrow |f\rangle $. As a consequence, the state (18) becomes
\begin{equation}
\frac{1}{\sqrt{2}}(|f\rangle |e\rangle +|e\rangle |f\rangle )|0\rangle
_{1}\left\vert 0\right\rangle _{2}\left\vert 0\right\rangle _{3}\left\vert
0\right\rangle _{4}.
\end{equation}%
Here and below, we assume $\Omega _{gf},\Omega _{ge}\gg g$ so that the
interaction between the qutrits and the cavities is negligible during the
application of the pulses.

The double NOON states are generated through the following $N$ steps of
operation:

Step 1: Let qutrit \textit{L} (\textit{R}) interact with the two cavities 1
and 2 (3 and 4) [Fig.~2(a)]. Based on Eqs. (11, 12, 14, 15), one can see that
after an interaction time $t_{1}=\pi /\left( 2\lambda \right) $, the state
(20) becomes
\begin{equation}
\frac{1}{\sqrt{2}}\left( |g\rangle |e\rangle |1\rangle _{1}|1\rangle
_{2}|0\rangle _{3}|0\rangle _{4}+|e\rangle |g\rangle |0\rangle _{1}|0\rangle
_{2}|1\rangle _{3}|1\rangle _{4}\right) ,
\end{equation}%
where a common phase factor $i$ is dropped off. Then, apply a classical
pulse of $\{\omega _{gf},\frac{\pi }{2\Omega _{gf}},-\frac{\pi }{2}\}$ to
the qutrits [Fig.~2(b)], resulting in the $|g\rangle \rightarrow |f\rangle $
for each qutrit. Thus, after the pulse, the state (21) changes to
\begin{equation}
\frac{1}{\sqrt{2}}\left( |f\rangle |e\rangle |1\rangle _{1}|1\rangle
_{2}|0\rangle _{3}|0\rangle _{4}+|e\rangle |f\rangle |0\rangle _{1}|0\rangle
_{2}|1\rangle _{3}|1\rangle _{4}\right) .
\end{equation}

Step $j$ ($j=2,3,...,N-1$): Repeat the manipulation of step 1 [Fig.~2(a),
Fig.~2(b)]. The time for each qutrit interacting with its two cavities is $%
t_{j}=\pi /\left( 2j\lambda \right) $ (i.e., half a Rabi oscillation).
According to Eq.~(11) and Eq.~(14), one can find that after an interaction
time $t_{j},$ the state $|f\rangle _{L}|j-1\rangle _{1}|j-1\rangle _{2}$ ($%
|f\rangle _{R}|j-1\rangle _{3}|j-1\rangle _{4}$) changes to $ie^{ij\lambda
t_{j}}|g\rangle _{L}|j\rangle _{1}|j\rangle _{2}$ ($ie^{ij\lambda
t_{j}}|g\rangle _{R}|j\rangle _{3}|j\rangle _{4}$) which further turns into $%
ie^{ij\lambda t_{j}}|f\rangle _{L}|j\rangle _{1}|j\rangle _{2}$ ($%
ie^{ij\lambda t_{j}}|f\rangle _{R}|j\rangle _{3}|j\rangle _{4}$) due to a
microwave pulse of $\{\omega _{gf},\frac{\pi }{2\Omega _{gf}},-\frac{\pi }{2}%
\}$ pumping the state $\left\vert g\right\rangle $ back to $\left\vert
f\right\rangle .$ Meanwhile, according to Eq. (12) and Eq. (15), both of the
states $|e\rangle _{L}|0\rangle _{3}|0\rangle _{4}$ ($|e\rangle
_{R}|0\rangle _{1}|0\rangle _{2}$) change to $e^{-i\lambda t_{j}}|e\rangle
_{L}|0\rangle _{3}|0\rangle _{4}$ ($e^{-i\lambda t_{j}}|e\rangle
_{R}|0\rangle _{1}|0\rangle _{2}$). Hence, one can easily verify that after
the operation of steps ($2,3,...,N-1$), the state (22) changes to
\begin{equation}
\frac{1}{\sqrt{2}}\left( |f\rangle |e\rangle |N-1\rangle _{1}|N-1\rangle
_{2}|0\rangle _{3}|0\rangle _{4}+|e\rangle |f\rangle |0\rangle _{1}|0\rangle
_{2}|N-1\rangle _{3}|N-1\rangle _{4}\right) ,
\end{equation}%
where a common phase factor $i^{N-2}e^{i\varphi }$ with $\varphi =\lambda
\sum\limits_{j=2}^{N-1}\left( j-1\right) t_{j}$ is dropped off.

Step $N$: Apply a classical pulse $\{\omega _{ge},\frac{\pi }{2\Omega _{ge}},%
\frac{\pi }{2}\}$ to each qutrit [Fig.~2(c)]. The interaction Hamiltonian is
\begin{equation}
H_{I,3}=\sum\limits_{j=L,R}(\Omega _{{ge}}e^{-i\pi /2}\sigma
_{ge,j}^{+}+h.c.).
\end{equation}%
It is easy to show that under this Hamiltonian, the applied pulses lead to
the transformation $|e\rangle \rightarrow |g\rangle $ for each qutrit. Thus,
the state (23) becomes
\begin{equation}
\frac{1}{\sqrt{2}}\left( |f\rangle |g\rangle |N-1\rangle _{1}|N-1\rangle
_{2}|0\rangle _{3}|0\rangle _{4}+\left\vert g\right\rangle |f\rangle
|0\rangle _{1}|0\rangle _{2}|N-1\rangle _{3}|N-1\rangle _{4}\right) ,
\end{equation}%
Meanwhile, let qutrit \textit{L} (\textit{R}) interact with the two cavities
1 and 2 (3 and 4) [Fig.~2(a)]. According to Eq.~(11) and Eq.~(14), one can
see that after an interaction time $t_{N}=\pi /\left( 2N\lambda \right) ,$
the state (25) changes to
\begin{equation}
\frac{1}{\sqrt{2}}\left( |N\rangle _{1}|N\rangle _{2}|0\rangle _{3}|0\rangle
_{4}+|0\rangle _{1}|0\rangle _{2}|N\rangle _{3}|N\rangle _{4}\right)
\left\vert g\right\rangle |g\rangle ,
\end{equation}%
where a common phase factor $ie^{i\left( N-1\right) \lambda t}$ is dropped
off.

Eq.~(26) shows that the four cavities (1, 2, 3, 4) are prepared in a double
NOON state and disentangled from the qutrits. The level spacings of the
qutrits need to be adjusted back to the original configuration after the
operation, so that each qutrit is decoupled from their own two cavities and
the central cavity, and thus the prepared double NOON state (26) remains
unchanged.

The above description shows that no adjustment of the cavity frequencies is
needed during the entire operation. The double-NOON-state generation
utilizes classical pulses with only two different frequencies (i.e., $\omega
=\omega _{gf},\omega _{ge}$), which are readily achieved in experiments.
Moreover, no measurement on the states of the coupler qutrits or the
cavities is required.

The total operational time (including the initial preparation for the Bell
state)\ is given by

\begin{equation}
t_{op}=\frac{\pi }{2\sqrt{2}\mu }+\sum\limits_{j=1}^{N}\frac{\pi \delta }{%
2jg^{2}}+\frac{N\pi }{2\Omega _{gf}}+\frac{\pi }{2\Omega _{ge}}+4t_{d},
\end{equation}%
where $t_{d}\sim 1-3$ ns is the typical time required for adjusting the
qutrit level spacings \cite{s7,a1}. To reduce decoherence from the qutrits
and the cavities, the operation time $t_{op}$ needs to be much shorter than
the energy relaxation time $T_{k,1}$ and dephasing time $T_{k,2}$ of the
level $\left\vert k\right\rangle $ ($k=f,e$). In principle, the $t_{op}$ can
be shortened by increasing the coupling constant $\mu $, the pulse Rabi
frequency $\Omega _{gf}$ and $\Omega _{ge}$, and by rapidly adjusting the
level spacings of the qutrits.

For cavity $k$\ ($k=1,2,3,4$), the lifetime of the cavity mode is given by $%
T_{cav}^{k}=\left( Q_{k}/\omega _{k}\right) /\overline{n}_{k},$\ where $%
Q_{k} $, $\omega _{k}$\ and $\overline{n}_{k}$\ are the (loaded) quality
factor, frequency, and the average photon number of cavity $k$,
respectively. For the four cavities here, the lifetime of entanglement of
the cavity modes is given by
\begin{equation}
T_{cav}\mathbf{=}\frac{1}{4}\min \mathbf{\{}T_{cav}^{1}\mathbf{,}%
T_{cav}^{2},T_{cav}^{3}\mathbf{,}T_{cav}^{4}\mathbf{\},}
\end{equation}%
which should be much longer than $t_{op},$ such that the effect of cavity
decay is negligible during the operation. It is noted that decoherence from
the central cavity can be neglected because the photon was populated in the
central cavity for a very short time. In principle, the $T_{cav}\gg t_{op}$
can be met by choosing cavities with a high quality factor.

The inter-cavity cross coupling between the two cavities on the left (right)
side is determined mostly by the coupling capacitances $c_{1}$ and $c_{2}$ ($%
c_{3}$ and $c_{4})\ $as well as the qutrit's self capacitance $C_{q}$,
because the\textbf{\ }field leakage through space\textbf{\ }is extremely low
for high-$Q$\ resonators as long as the inter-cavity distance is much
greater than transverse dimension of the cavities - a condition easily met
in experiments for the two resonators \cite{s21}. As our numerical
simulation shows below [see Fig.~4(a)], the operational fidelity is insensitive
to the crosstalk of cavities $1$ and $2$, when the detuning
between the frequencies of cavities $1$ and $2$ is much larger than
the inter-cavity coupling constant $g_{12}$ between cavities $1$ and $2$.
The same holds for cavities $3$ and $4$). It is noted that the inter-cavity
crosstalk between the side cavities and the central cavity can be neglected by
adjusting the central cavity frequency such that the central cavity frequency
is highly detuned from the side-cavity frequencies.

\begin{center}
\textbf{V. POSSIBLE EXPERIMENTAL IMPLEMENTATION}
\end{center}

In this section, we discuss the fidelity for generating double NOON
states with photon number $N\leq 10$. In the numerical calculations, the
preparing of the initial Bell state will not be considered, since it is
extremely fast due to using the resonant interaction only.

We will consider inter-cavity crosstalks of cavities on both sides, which
can be described as $\varepsilon =g_{12}\left( e^{i\Delta
t}a_{1}a_{2}^{+}+h.c.\right) +g_{34}\left( e^{i\Delta
t}a_{3}a_{4}^{+}+h.c.\right) $, where $g_{12}$ ($g_{34}$) is the coupling
strength of cavities $1$ and $2$ ($3$ and $4$), and we have assumed the
detuning between cavity frequencies $\omega _{c_{1}}$ and $\omega _{c_{2}}$ (%
$\omega _{c_{3}}$ and $\omega _{c_{4}}$) $\Delta =\omega _{c_{2}}-\omega
_{c_{1}}=\omega _{c_{4}}-\omega _{c_{3}}$ for simplicity. We will also
consider the unwanted qutrit-cavity interactions and inter-cavity crosstalks
during the application of pulses. So the Hamiltonians $H_{I,1},$ $H_{I,2},$
and $H_{I,3}$ adopted in the double NOON state generation should be modified
to $\widetilde{H}_{I,1}=H_{I,1}+\varepsilon ,$ $\widetilde{H}_{I,2}=H_{I,2}+%
\widetilde{H}_{I,1},$ and $\widetilde{H}_{I,3}=H_{I,3}+\widetilde{H}_{I,1}$,
respectively.

By considering dissipation and dephasing, the evolving of the system is
determined by the master equation
\begin{eqnarray}
\frac{d\rho }{dt} &=&-i\left[ \widetilde{H}_{I,k},\rho \right]
+\sum_{l=1}^{4}\kappa _{a_{l}}\mathcal{L}\left[ a_{l}\right]   \notag \\
&&+\sum_{j=L,R}\gamma _{ef,j}\mathcal{L}\left[ \sigma _{ef,j}^{-}\right]
+\gamma _{gf,j}\mathcal{L}\left[ \sigma _{gf,j}^{-}\right] +\gamma _{ge,j}%
\mathcal{L}\left[ \sigma _{ge,j}^{-}\right]   \notag \\
&&+\sum_{j=L,R}\gamma _{f\varphi ,j}\left( \sigma _{ff,j}\rho \sigma
_{ff,j}-\sigma _{ff,j}\rho /2-\rho \sigma _{ff,j}/2\right)   \notag \\
&&+\sum_{j=L,R}\gamma _{e\varphi ,j}\left( \sigma _{ee,j}\rho \sigma
_{ee,j}-\sigma _{ee,j}\rho /2-\rho \sigma _{ee,j}/2\right) ,
\end{eqnarray}%
where $\widetilde{H}_{I,k}$ (with $k=1,2,3$) are the modified $\widetilde{H}%
_{I,1},$ $\widetilde{H}_{I,2},$ and $\widetilde{H}_{I,3}$, $\mathcal{L}\left[
\Lambda \right] =\Lambda \rho \Lambda ^{+}-\Lambda ^{+}\Lambda \rho /2-\rho
\Lambda ^{+}\Lambda /2$ (with $\Lambda =a_{l},\sigma _{ef,j}^{-},\sigma
_{gf,j}^{-},\sigma _{ge,j}^{-})$,\ $\sigma _{ef,j}^{-}=\left\vert
e\right\rangle _{j}\left\langle f\right\vert ,$ $\sigma
_{gf,j}^{-}=\left\vert g\right\rangle _{j}\left\langle f\right\vert ,$ $%
\sigma _{ge,j}^{-}=\left\vert g\right\rangle _{j}\left\langle e\right\vert ,$
$\sigma _{ff,j}=\left\vert f\right\rangle _{j}\left\langle f\right\vert $,
and $\sigma _{ee,j}=\left\vert e\right\rangle _{j}\left\langle e\right\vert ;
$ $\kappa _{a_{l}}$ is the decay rate of cavity $l$ ($l=1,2,3,4$);\ $\gamma
_{ef,j}$ ($\gamma _{gf,j}$) is the energy relaxation rate for the level $%
\left\vert f\right\rangle $\ associated with the decay path $\left\vert
f\right\rangle \rightarrow \left\vert e\right\rangle $ ($\left\vert
f\right\rangle \rightarrow \left\vert g\right\rangle $) of qutrit $j$; $%
\gamma _{ge,j}$\ is the energy relaxation rate of the level $\left\vert
e\right\rangle ;$and $\gamma _{f\varphi ,j}$ ($\gamma _{e\varphi ,j}$) is
the dephasing rate of the level $\left\vert f\right\rangle $ ($\left\vert
e\right\rangle $) of qutrit $j$ ($j=L,R$)\textbf{. }The fidelity of the
whole operation is given by $\mathcal{F}=\sqrt{\left\langle \psi
_{id}\right\vert \rho \left\vert \psi _{id}\right\rangle },$ where $%
\left\vert \psi _{id}\right\rangle $ is the ideal output state given in
Eq.~(26), while $\rho $ is the final density matrix obtained by numerically
solving the master equation. For numerical calculations, we here use the
QUTIP software \cite{s211,s212}, which is an open-source software for
simulating the dynamics of open quantum systems.

We now numerically calculate the fidelity. For flux qutrits, the transition
frequency between two neighbor levels can be made to be $1-20$ GHz. As an
example, we choose the frequencies of qutrits as $\omega _{ef}/2\pi \sim 7.5$
GHz and $\omega _{ge}/2\pi \sim 5.0$ GHz, and the detuning $\delta /2\pi =1.0
$ GHz. As a result, we have $\omega _{c_{1}}/2\pi ,\omega _{c_{3}}/2\pi \sim
4.0$ GHz, and $\omega _{c_{2}}/2\pi ,\omega _{c_{4}}/2\pi \sim 8.5$ GHz, and
thus $\Delta /2\pi \sim 4.5$ GHz. For simplicity, we set $\Omega _{{ge}%
}=\Omega _{{gf}}=\Omega ,$ which can be readily achieved by adjusting the
pulse intensity. Other parameters used in the numerical simulations are: (i)
$\gamma _{e\varphi ,j}^{-1}=5$ $\mu $s$,$ $\gamma _{f\varphi ,j}^{-1}=5$ $%
\mu $s; (ii) $\gamma _{ge,j}^{-1}=10$ $\mu $s, $\gamma _{ef,j}^{-1}=5$ $\mu $%
s, $\gamma _{gf,j}^{-1}=20$ $\mu $s \cite{s16,s17,s18,s19,s20}, and (iii) $%
\kappa _{a_{l}}^{-1}=20$ $\mu $s ($l=1,2,3,4$). For the cavity frequencies
and $\kappa _{a_{l}}^{-1}$ used here, the required quality factors for the
four cavities are $Q_{1}=Q_{3}\sim 5.0\times 10^{5}$ and $Q_{2}=Q_{4}\sim
1.0\times 10^{6},$ which are readily available in experiments \cite{s22,s23}.

\begin{figure}[tbp]
\begin{center}
\includegraphics[bb=0 74 1216 1000, width=14 cm, clip]{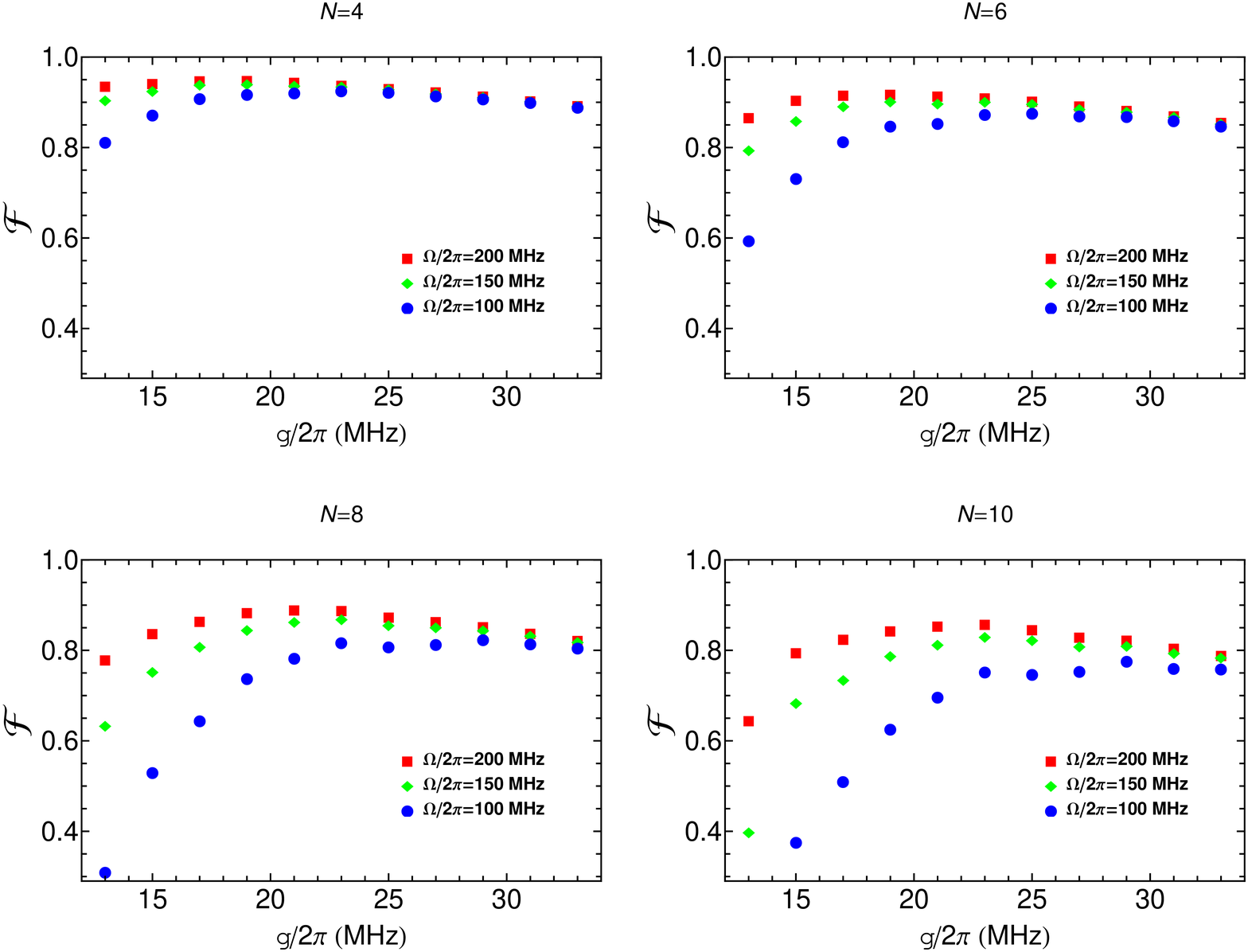} \vspace*{%
-0.08in}
\end{center}
\caption{(Color online) Fidelities versus $g/2\protect\pi$ for
double NOON state generation with photon numbers of $N=4, 6, 8 ,10$.
For each $N$, $\Omega/2\protect\pi=100, 150, 200$ MHz \protect\cite{s15} are
considered. The figure was plotted by assuming $g_1=g_2=g$ and $g_{12}=g_{34}=0$.}
\label{fig:3}
\end{figure}

\begin{figure}[tbp]
\begin{center}
\includegraphics[bb=0 0 875 248, width=16 cm, clip]{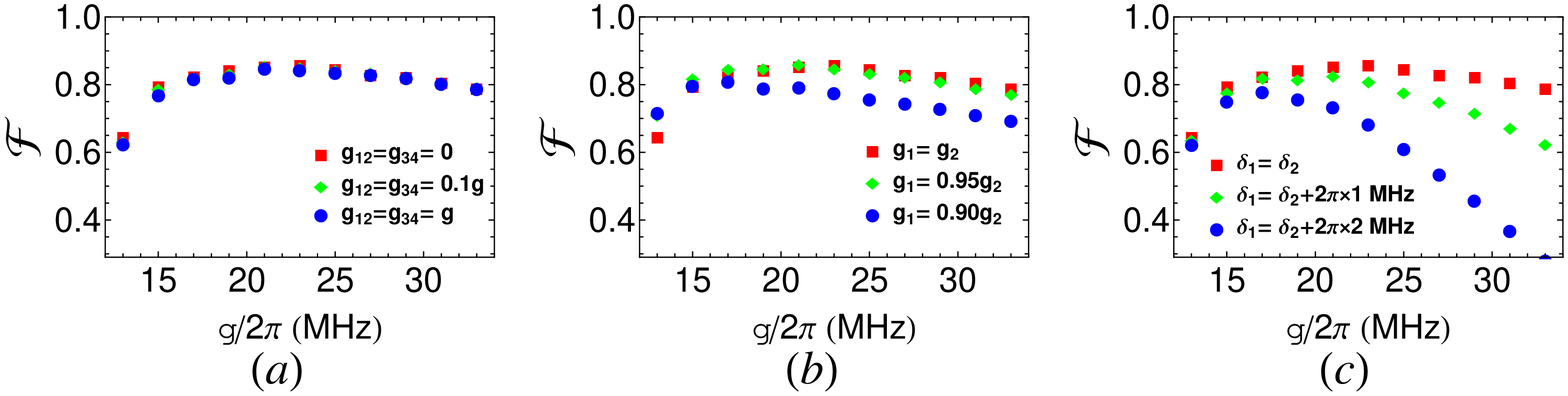} \vspace*{%
-0.08in}
\end{center}
\caption{(Color online) Fidelities versus $g/2\protect\pi$ with
photon numbers of $N=10$. (a) Cases for
$g_{12}=g_{34}=0$, $g_{12}=g_{34}=0.1g$, and $g_{12}=g_{34}=g$. (b)
Cases for $g_{1}=g_{2}$, $g_{1}=0.95g_{2}$, and $g_{1}=0.9g_{2}$
(setting $g_{2}=g$). (c) Cases for $\delta_1$=$\delta_2$,
$\delta_1=\delta_2+2\pi\times1$ MHz, and $\delta_1=\delta_2+2\pi\times2$ MHz (setting $\delta_2=\delta$).
In (a), (b), and (c), other parameters used in the numerical simulations are the same as those used in Fig.~3.}
\label{fig:4}
\end{figure}

In Fig.~3, fidelities versus $g/2\pi $ are plotted respectively for $%
N=4,6,8,10$ and $\Omega /2\pi =100,150,200$ MHz \cite{s15}, with $%
g_{1}=g_{2}=g$ and $g_{12}=g_{34}=0$. For $\Omega /2\pi =150$ MHz, the best
fidelity and the corresponding optimal $g/2\pi $ are: $\{0.941,19$ MHz$\}\
(N=4)$; $\{0.903,19$ MHz$\}\ (N=6)$; $\{0.870,23$ MHz$\}\ (N=8)$; $%
\{0.831,23 $ MHz$\}\ (N=10)$. If $\Omega /2\pi $ is increased to $200$ MHz,
the fidelity will be improved, as shown in Fig.~3.

To see the effect of the inter-cavity crosstalk and the parameter
inhomogeneity on the fidelity, we consider: (i) $g_{12}=g_{34}=0$, $%
g_{12}=g_{34}=0.1g,$ and $g_{12}=g_{34}=g$ in Fig.~4(a), (ii) $g_{1}=g_{2},$
$g_{1}=0.95g_{2},$ and $g_{1}=0.9g_{2}$ in Fig.~4(b), and (iii) $\delta
_{1}=\delta _{2},$ $\delta _{1}=\delta _{2}+2\pi \times 1$ MHz, and $\delta
_{1}=\delta _{2}+2\pi \times 2$ MHz in Fig.~4(c). Fig.~4 is plotted for $N=10
$ and $\Omega /2\pi =200$ MHz. Fig.~4(a) shows that the effect of the inter-cavity crosstalk
on the fidelity is negligible. This is due to the large detuning between
the cavity frequencies ($\Delta /2\pi \sim 4.5$ GHz), compared to $%
g_{12}$ and $g_{34}$. Fig.~4(b) shows that compared to the homogeneous case, a small difference
between $g_{1}$ and $g_{2}$ only leads to a small change of fidelity. In
contrast, Fig.~4(c) shows that the fidelity decreases fast as the
difference between $\delta _{1}$ and $\delta _{2}$ increases.

\begin{center}
\textbf{VI. CONCLUSIONS}
\end{center}

We have presented an approach for generating the double NOON states of
photons in circuit QED. We believe that this work is of interest because our
work is the first to demonstrate that the double NOON states of photons can
be generated in circuit QED with only $N+2$ operational steps. The numerical
simulations show that high-fidelity generation of the double NOON states
with $N\leqslant 10$ even for the imperfect devices is feasible with
present-day circuit QED technique. The double NOON states are entangled
states on more parties and thus different from the traditional NOON states.
Compared to the NOON states, the double NOON states may achieve the same
error in phase measurement using the same number of photons while requiring
much less operational steps to prepare.

\begin{center}
\textbf{ACKNOWLEDGMENTS}
\end{center}

C.P. Yang and Q.P. Su were supported in part by the Ministry of Science and
Technology of China under Grant No. 2016YFA0301802, the National Natural
Science Foundation of China under Grant Nos 11504075, 11074062, and
11374083, and the Zhejiang Natural Science Foundation under Grant No.
LZ13A040002. J.M. Liu was supported in part by the National Natural Science
Foundation of China under Grant Nos 11174081 and 11134003, the National
Basic Research Program of China under Grant No. 2012CB821302, and the
Natural Science Foundation of Shanghai under Grant No. 16ZR1448300. This
work was also supported by the funds from Hangzhou City for the
Hangzhou-City Quantum Information and Quantum Optics Innovation Research
Team.

\begin{center}
\textbf{APPENDIX A}
\end{center}

\begin{figure}[tbp]
\begin{center}
\includegraphics[bb=39 173 550 666, width=7.5 cm, clip]{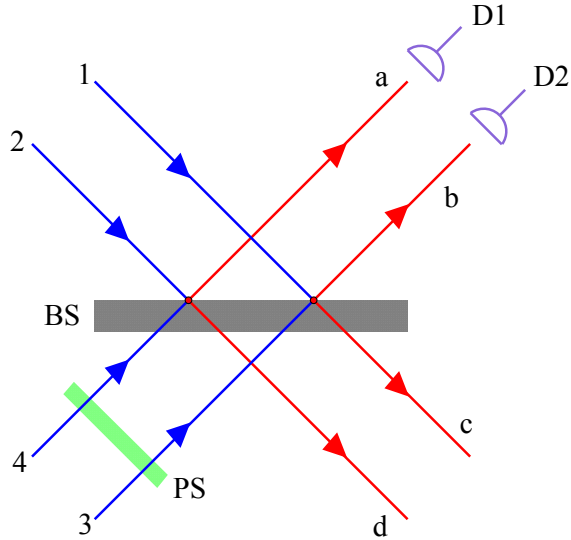} \vspace*{%
-0.08in}
\end{center}
\caption{(Color online) Setup for phase measurement by using a double
NOON states. Initially, photons in four modes ($1,2,3,4$) are in a
double NOON state $(|NN00\rangle _{1234}+|00NN\rangle
_{1234})/\sqrt{2}$. Each Photon in mode $3$ ($4$) experiences an extra
phase shift $\phi$ compared to each photon in mode $1$ ($2$), which is
induced by a phase shifter (PS). Here, BS represents a beam splitter, D1 and D2 represent photon detectors.
The probability for coincidence measurement of $N$ photons at mode a and $N$ photons at mode b is
$P=\eta [1+\cos(2N\phi)]/2$ with $0<\eta=2^{1-2N}<1$ for $N\neq 0$.}\label{fig:5}
\end{figure}

Here we show how to implement the measurement of
$A_{DN}=(|NN00\rangle _{1234}+|00NN\rangle _{1234})(\langle
NN00|_{1234}+\langle 00NN|_{1234})/2$. A simple setup for this measurement is shown in
Fig.~5. Suppose photons in four modes ($1,2,3,4$) are initially in a
double NOON state $(|NN00\rangle _{1234}+|00NN\rangle
_{1234})/\sqrt{2}$. The evolution of state of photons in four modes
can be expressed as follows:
\begin{eqnarray}
&&\frac{1}{\sqrt{2}}(|NN00\rangle _{1234}+|00NN\rangle _{1234})\nonumber\\
&=&\frac{1}{\sqrt{2}N!}[(A_1^+A_2^+)^{N}+(A_3^+A_4^+)^{N}]\cdot|0\rangle\nonumber\\
&\underrightarrow{PS}&\frac{1}{\sqrt{2}N!}[(A_1^+A_2^+)^{N}+e^{i2N\phi}(A_3^+A_4^+)^{N}]\cdot|0\rangle\nonumber\\
&\underrightarrow{BS}&\frac{1}{2^N\sqrt{2}N!}[(A_b^++A_c^+)^{N}(A_a^++A_d^+)^{N}+e^{i2N\phi}(A_b^+-A_c^+)^{N}(A_a^+-A_d^+)^{N}]\cdot|0\rangle\nonumber\\
&=&\frac{1}{2^N\sqrt{2}N!}(1+e^{i2N\phi})(A_a^+A_b^+)^N\cdot|0\rangle+...\nonumber\\
&=&\frac{1}{2^N\sqrt{2}}(1+e^{i2N\phi})|NN00\rangle
_{abcd}+...~,\nonumber
\end{eqnarray}
where $A_i^+$ is the photon creation operator of mode $i$
($i=1,2,3,4,a,b,c,d$). The first term of the last line indicates that the probability of
coincidence measurement of $N$ photons in mode a and $N$ photons at mode b is
$P=\eta[1+\cos(2N\phi)]/2$, which is just the expectation of
$A_{DN}$ times an additional constant $\eta=2^{1-2N}$. Note that the
appearance of $\eta$ here does not affect the error of phase
measurement $\Delta\phi$.


\begin{thebibliography}{99}
\bibitem{s01} J. Q. You and F. Nori, \textquotedblleft Superconducting
circuits and quantum information", Phys. Today \textbf{58}(11), 42 (2005).

\bibitem{s02} J. Q. You and F. Nori, ``Atomic physics and quantum optics
using superconducting circuits'', Nature \textbf{474}, 589 (2011).

\bibitem{s03} I. Buluta, S. Ashhab, and F. Nori, \textquotedblleft Natural
and artificial atoms for quantum computation\textquotedblright , Rep. Prog.
Phys. \textbf{74}, 104401 (2011).

\bibitem{s04} Z. L. Xiang, S. Ashhab, J. Q. You, and F. Nori, ``Hybrid
quantum circuits: Superconducting circuits interacting with other quantum
systems'', Rev. Mod. Phys. \textbf{85}, 623 (2013).

\bibitem{s05} R. Barends, J. Kelly, A. Megrant, D. Sank, E. Jeffrey, Y.
Chen, Y. Yin, B. Chiaro, J. Y. Mutus, C. Neill, P. J. J. O'Malley, P.
Roushan, J. Wenner, T. C. White, A. N. Cleland, and J. M. Martinis,
\textquotedblleft Coherent Josephson qubit suitable for scalable quantum
integrated circuits\textquotedblright , Phys. Rev. Lett. \textbf{111},
080502 (2013).

\bibitem{s06} M. Neeley, M. Ansmann, R. C. Bialczak, M. Hofheinz, N. Katzl,
E. Lucero, A. O'Connell, H. Wang, A. N. Cleland, and J. M. Martinis,
``Process tomography of quantum memory in a Josephson-phase qubit coupled to
a two-level state'', Nat. Phys. \textbf{4}, 523 (2008).

\bibitem{s07} P. J. Leek, S. Filipp, P. Maurer, M. Baur, R. Bianchetti, J.
M. Fink, M. Goppl, L. Steffen, and A. Wallraff, ``Using sideband transitions
for two-qubit operations in superconducting circuits'', Phys. Rev. B \textbf{%
79}, 180511(R) (2009).

\bibitem{s08} J. D. Strand, M. Ware, F. Beaudoin, T. A. Ohki, B. R. Johnson,
A. Blais, and B. L. T. Plourde, ``First-order sideband transitions with
flux-driven asymmetric transmon qubits'', Phys. Rev. B \textbf{87},
220505(R) (2013).

\bibitem{s081} J. K. Chow, J. M. Gambetta, A. D. Crcoles, S. T. Merkel, J.
A. Smolin, C. Rigetti, S. Poletto, G. A. Keefe, M. B. Rothwell, J. R. Rozen,
M. B. Ketchen, and M. Steffen, ``Universal Quantum Gate Set Approaching
Fault-Tolerant Thresholds with Superconducting Qubits'', Phys. Rev. Lett.
\textbf{109}, 060501 (2012).

\bibitem{s082} J. B. Chang, M. R. Vissers, A. D. Corcoles, M. Sandberg, J.
Gao, D. W. Abraham, J. M. Chow, J. M. Gambetta, M. B. Rothwell, G. A. Keefe,
M. Steffen, and D. P. Pappas, ``Improved superconducting qubit coherence
using titanium nitride'', Appl. Phys. Lett. \textbf{103}, 012602 (2013).

\bibitem{s083} R. Barends, J. Kelly, A. Megrant, D. Sank, E. Jeffrey, Y.
Chen, Y. Yin, B. Chiaro, J. Y. Mutus, C. Neill, P. J. J. O'Malley, P.
Roushan, J. Wenner, T. C. White, A. N. Cleland, and J. M. Martinis,
``Coherent Josephson Qubit Suitable for Scalable Quantum Integrated
Circuits'', Phys. Rev. Lett. \textbf{111} , 080502 (2013).

\bibitem{s084} J. M. Chow, J. M. Gambetta, E. Magesan, D. W. Abraham, A. W.
Cross, B. R. Johnson, N. A. Masluk, C. A. Ryan, J. A. Smolin, S. J.
Srinivasan, and M. Steffen, ``Implementing a strand of a scalable
fault-tolerant quantum computing fabric'', Nat. Commun. \textbf{5}, 4015
(2014).

\bibitem{s085} Y. Chen, C. Neill, P. Roushan, N. Leung, M. Fang, R. Barends,
J. Kelly, B. Campbell, Z. Chen, B. Chiaro, A. Dunsworth, E. Jeffrey, A.
Megrant, J. Y. Mutus, P. J. J. O'Malley, C. M. Quintana, D. Sank, A.
Vainsencher, J. Wenner, T. C. White, M. R. Geller, A. N. Cleland, and J. M.
Martinis, ``Qubit Architecture with High Coherence and Fast Tunable
Coupling'', Phys. Rev. Lett. \textbf{113}, 220502 (2014).

\bibitem{s086} M. Stern, G. Catelani, Y. Kubo, C. Grezes, A. Bienfait, D.
Vion, D. Esteve, and P. Bertet, ``Flux Qubits with Long Coherence Times for
Hybrid Quantum Circuits'', Phys. Rev. Lett. \textbf{113}, 123601 (2014)

\bibitem{s09} A. Wallraff, D. I. Schuster, A. Blais, L. Frunzio, R. S.
Huang, J. Majer, S. Kumar, S. M. Girvin, and R. J. Schoelkopf,
\textquotedblleft Strong coupling of a single photon to a superconducting
qubit using circuit quantum electrodynamics\textquotedblright , Nature
\textbf{431}, 162 (2004).

\bibitem{s00} T. Niemczyk, F. Deppe, H. Huebl, E. P. Menzel, F. Hocke, M. J.
Schwarz, J. J. Garcia-Ripoll, D. Zueco, T. Hummer, E. Solano, A. Marx, and
R. Gross, ``Circuit quantum electrodynamics in the ultrastrong-coupling
regime'', Nat. Phys. \textbf{6}, 772 (2010).

\bibitem{s1} A. N. Boto, P. Kok, D. S. Abrams, S. L. Braunstein, C. P.
Williams, and J. P. Dowling, \textquotedblleft Quantum interferometric
optical lithography: Exploiting entanglement to beat the diffraction limit",
Phys. Rev. Lett. \textbf{85}, 2733 (2000).

\bibitem{s4} M. D'Angelo, M. V. Chekhova, and Y. Shih, ``Two-photon
Diffraction and Quantum Lithography", Phys. Rev. Lett. \textbf{87}, 013602
(2001).

\bibitem{s2} P. Kok, H. Lee, and J. P. Dowling, \textquotedblleft Creation
of large-photon-number path entanglement conditioned on photodetection",
Phys. Rev. A \textbf{65}, 052104 (2002).

\bibitem{s201} M. W. Mitchell, J. S. Lundeen, and A. M. Steinberg,
\textquotedblleft Super-resolving phase measurements with a multi-photon
entangled state\textquotedblright , Nature 429, 161 (2004).

\bibitem{s202} M. Muller, H. Vural, and P. Michler, ``Phase super-resolution
with N00N states generated by on demand single-photon sources'',
arXiv:1603.00906

\bibitem{s3} J. Joo, W. J. Munro, and T. P. Spiller, ``Quantum metrology
with entangled coherent states", Phys. Rev. Lett. \textbf{107}, 083601
(2011).

\bibitem{s31} C. H. Bennett and B. D. DiVincenzo, ``Quantum information and
computation'', Nature 404, 247 (2000).

\bibitem{s41} W. Heitler, ``The Quantum Theory of Radiation'', 3rd ed.
(Oxford University Press, London, 1954).

\bibitem{s8} F. W. Strauch, K. Jacobs and R. W. Simmonds, ``Arbitrary
control of entanglement between two superconducting resonators", Phys. Rev.
Lett. \textbf{105}, 050501 (2010).

\bibitem{s9} S. T. Merkel and F. K. Wilhelm, \textquotedblleft Generation
and detection of NOON states in superconducting circuits", New J. Phys.
\textbf{12}, 093036 (2010).

\bibitem{s5} H. Wang, M. Mariantoni, R. C. Bialczak, M. Lenander, E. Lucero,
M. Neeley, A. O'Connell, D. Sank, M. Weides, J. Wenner, T. Yamamoto, Y. Yin,
J. Zhao, J. M. Martinis, and A. N. Cleland, ``Deterministic entanglement of
photons in two superconducting microwave resonators", Phys. Rev. Lett.
\textbf{106}, 060401 (2011).

%\bibitem{s6} F. W. Strauch, ``All-Resonant Control of Superconducting
%Resonators", Phys. Rev. Lett. \textbf{109}, 210501 (2012).

\bibitem{s7} Q. P. Su, C. P. Yang, and S. B. Zheng, ``Fast and simple scheme
for generating NOON states of photons in circuit QED", Scientific Reports
\textbf{4}, 3898 (2014).

\bibitem{s13} S. J. Xiong, Z. Sun, J. M. Liu, T. Liu, and C. P. Yang,
\textquotedblleft Efficient scheme for generation of photonic NOON states in
circuit QED", Optics Letters \textbf{40}, 2221 (2015).

\bibitem{s14} D. F. V. James, \textquotedblleft Effective Hamiltoniant
theory and its applications in quantum information", arXiv:0706.1090.

\bibitem{s11} J. Clarke and F. K. Wilhelm, ``Superconducting quantum bits",
Nature \textbf{453}, 1031 (2008).

%\bibitem{a1} Q.P.Su, C.P. Yang, and S.B.Zheng, ``Fast and simple scheme for generating NOON states of photons in circuit QED'', Sci. Rep. 4, 3898 (2014);

\bibitem{a1} Y. Yu and S. Han, Private communication (2015).

\bibitem{s15} M. Baur, S. Filipp, R. Bianchetti, J. M. Fink, M. Gpl, L.
Steffen, P. J. Leek, A. Blais, and A. Wallraff, \textquotedblleft
Measurement of Autler-Townes and mollow transitions in a strongly driven
superconducting qubit\textquotedblright , Phys. Rev. Lett. \textbf{102},
243602 (2009).

\bibitem{s16} F. Yan, S. Gustavsson, A. Kamal, J. Birenbaum, A. P. Sears, D.
Hover, T. J. Gudmundsen, D. Rosenberg, G. Samach, and S. Weber \textit{%
et al.}, \textquotedblleft The Flux qubit revisited to enhance coherence and reproducibility\textquotedblright ,
Nat. Commun. \textbf{7}, 12964 (2016).

\bibitem{s17} J. Q. You, X. Hu, S. Ashhab, and F. Nori, ``Low-decoherence
flux qubit'', Phys. Rev. B \textbf{75}, 140515(R) (2007).

\bibitem{s18} M. J. Peterer, S. J. Bader, X. Jin, F. Yan, A. Kamal, T. J.
Gudmundsen, P. J. Leek, T. P. Orlando, W. D. Oliver, and S. Gustavsson,
``Coherence and decay of higher energy levels of a superconducting transmon
qubit'', Phys. Rev. Lett. \textbf{114}, 010501 (2015).

\bibitem{s19} C. Rigetti, J. M. Gambetta, S. Poletto, B. L. T. Plourde, J.
M. Chow, A. D. C\'{o}rcoles, J. A. Smolin, S. T. Merkel, J. R. Rozen, and G.
A. Keefe \textit{et al.}, ``Superconducting qubit in a waveguide cavity with
a coherence time approaching 0.1 ms'', Phys. Rev. B \textbf{86}, 100506(R)
(2012).

\bibitem{s20} I. M. Pop, K. Geerlings, G. Catelani, R. J. Schoelkopf, L. I.
Glazman, and M. H. Devoret, ``Coherent suppression of electromagnetic
dissipation due to superconducting quasiparticles'', Nature \textbf{508},
369 (2014).

\bibitem{s21} C. P. Yang, Q. P. Su, and S. Han, ``Generation of
Greenberger-Horne-Zeilinger entangled states of photons in multiple cavities
via a superconducting qutrit or an atom through resonant interaction'',
Phys. Rev. A \textbf{86}, 022329 (2012).

\bibitem{s211} J. R. Johansson, P. D. Nation, and F. Nori, QuTiP: An
opensource Python framework for the dynamics of open quantum systems,
Comput. Phys. Commun. 183, 1760 (2012).

\bibitem{s212} J. R. Johansson, P. D. Nation, and F. Nori, QuTiP 2: A Python
framework for the dynamics of open quantum systems, Comput. Phys. Commun.
184, 1234 (2013).

\bibitem{s22} W. Chen, D. A. Bennett, V. Patel, and J. E. Lukens,
``Substrate and process dependent losses in superconducting thin film
resonators'', Sci. Technol. \textbf{21} , 075013 (2008).

\bibitem{s23} P. J. Leek, M. Baur, J. M. Fink, R. Bianchetti, L. Steffen, S.
Filipp, and A. Wallraff, ``Cavity quantum electrodynamics with separate
photon storage and qubit readout modes'', Phys. Rev. Lett. \textbf{104},
100504 (2010).
\end{thebibliography}
\end{document}